\documentclass[10pt]{article}
\usepackage{amssymb}
\newcommand{\referencek}{{\mathsf k}}
\newcommand{\referencel}{{\mathsf l}}
\newcommand{\referenceK}{{\mathsf K}}
\newcommand{\referencesigma}{{\varsigma}}
\newcommand{\referenceSigma}{{\mathsf \Sigma}}
\newcommand{\referenceN}{{\mathsf N}}
\newcommand{\referenceM}{{\mathsf M}}
\begin{document}
\begin{center}
{\bf \large Lightcone reference for total 
gravitational energy${}^{*}$}\\[6mm] 
Stephen R.~Lau${}^{\dagger}$
\\[4mm]
Theoretical Astrophysics \& Relativity Group\\
Department of Physics \& Astronomy, 
University of North Carolina\\
CB\# 3255 Phillips Hall, Chapel Hill, 
NC 27599-3255 USA\\[2mm]
and\\[2mm]
Applied Mathematics Group\\ 
Department of Mathematics,
University of North Carolina\\
CB\# 3250 Phillips Hall, Chapel Hill, 
NC 27599-3250 USA\\[6mm]
{\bf Abstract}
\end{center}
We give an explicit expression for gravitational energy, 
written solely in terms of physical spacetime geometry,
which in suitable limits agrees with the total 
Arnowitt-Deser-Misner and Trautman-Bondi-Sachs energies
for asymptotically flat spacetimes and with the Abbot-Deser
energy for asymptotically anti-de Sitter spacetimes. Our 
expression is a boundary value of the standard 
gravitational Hamiltonian. Moreover, although it stands
alone as such, we derive the expression by picking 
the zero-point of energy via a ``lightcone reference.''
\vfill
\noindent \underline{Chapel Hill, 10 March 1999}

\noindent ${}^{*}$ {\tt gr-qc/yyymmnn}, TAR-067-UNC

\noindent ${}^{\dagger}${\tt lau@math.unc.edu}
\newpage

\section*{Introduction}
It is satisfying to learn that each of the accepted 
notions of total gravitational energy can be 
expressed as a limit of (the on-shell value of) the 
standard gravitational Hamiltonian $H$. Indeed, i.~the 
Arnowitt-Deser-Misner ({\sc adm}) energy \cite{ADM} 
associated with spatial infinity for asymptotically flat 
spacetimes, ii.~the Trautman-Bondi-Sachs ({\sc tbs}) 
energy \cite{TBS} associated with null infinity for 
asymptotically flat spacetimes, and iii.~the 
Abbott-Deser ({\sc ab}) energy \cite{AD} 
associated with infinity\footnote{Null infinity 
coincides with spatial infinity for asymptotically anti-de 
Sitter solutions.\cite{HawkingEllis}} for asymptotically 
anti-de Sitter spacetimes can all be written as (a limit 
of) the following Hamiltonian boundary value\footnote{As 
written by Brown and York.\cite{BrownYork}}:
\begin{equation}
      \left. H\right|_{B} = 
      (8\pi)^{-1}\int_{B} 
      {\rm d}^{2}x \sqrt{\sigma} 
      N (k - k|^{\rm ref})
\label{Hvalue}
\end{equation}
Here $B$ is a large nearly spherical two-surface 
with coordinate radius $r$, tending
to an infinite-area round sphere as $r \rightarrow 
\infty$. $N$ is a smearing (lapse) function, which
is unity for the asymptotically flat scenarios and 
grows like $r$ in the asymptotically anti-de Sitter
case, and ${\rm d}^{2}x\sqrt{\sigma}$ is the proper 
$B$ area element. $B$ is embedded 
in a three-slice $\Sigma$ of the ambient physical 
spacetime $M$, and $k$ is the mean curvature of 
$B$ as embedded in $\Sigma$. 
Usually, $k|^{\rm ref} = k|^{\rm ref}(\sigma_{ab})$ 
denotes an arbitrary function of the intrinsic $B$ 
metric $\sigma_{ab}$. However, to ensure that 
(\ref{Hvalue}) gives rise to the correct asymptotic
energies as $r \rightarrow \infty$, we define it 
in this paragraph via the following prescription.
\begin{itemize}
\item[$\bullet$]{Embed $B$ 
isometrically in a static-time slice 
$\referenceSigma$ of a reference spacetime 
$\referenceM$, which will either be Minkowski 
spacetime or an anti-de Sitter 
spacetime. For example, when defining 
the {\sc adm} energy, embed $B$ isometrically in 
Euclidean three-space $E^{3}$.}
\item[$\bullet$]{Compute 
$k|^{\rm ref}$ as the mean curvature of 
$B$ as embedded in $\referenceSigma$.}
\end{itemize} 
Physically, the proper surface integral of 
$(8\pi)^{-1} N k|^{\rm ref}$ is set as the energy
zero-point, and (\ref{Hvalue}) expresses the energy
of $\Sigma$ {\em relative} to the energy of 
$\referenceSigma$. This {\em choice} for the 
zero-point determines the correct asymptotic 
energies.\footnote{The {\sc adm} correspondence
was noted first in Refs.~\cite{BrownYork}, while the
{\sc tbs} correspondence was shown in Ref.~\cite{BLY1}.
For asymptotically anti-de Sitter spacetimes,
that the surface integral (\ref{Hvalue}) defines
the correct ``energy'' (actually ``conserved mass'' 
is a better term in this context) at infinity was
shown first in Ref.~\cite{Brown}. Ref.~\cite{HawkingHorowitz}
spells out the relationship between the integral 
(\ref{Hvalue}) and the original definition of Abbott
and Deser.} However, 
by following this prescription, we only define the 
surface integral (\ref{Hvalue}) {\em implicitly}. 
Indeed, obtaining a closed-form expression for 
$k|^{\rm ref}$ is tantamount to actually solving 
a stubborn embedding problem. For all but simple 
cases this is a hopeless proposition.

In this brief report, we show that the explicit 
alternative
\begin{equation}
\left. H\right|_B = (8\pi)^{-1}\int_B
{\rm d}^{2}x \sqrt{\sigma} N (k + 
\sqrt{2{\cal R} + 4/\ell^{2}})
\label{newHvalue}
\end{equation}
to (\ref{Hvalue}) also determines the correct 
asymptotic energies. Here ${\cal R}$ is the $B$ 
Ricci scalar, and $\ell$ is the radius of curvature
for an asymptotically anti-de Sitter solution, related
to the (negative) cosmological constant $\Lambda$ by
$\ell = (-3/\Lambda)^{1/2}$. One must take $\ell 
\rightarrow \infty$ for asymptotically flat 
spacetimes, in which case the second term under the 
radical vanishes. Now setting
\begin{equation}
      k|^{\rm ref} = 
      -\sqrt{2{\cal R} + 4/\ell^{2}}{\,},
\label{newreference}
\end{equation} 
we pick a different zero-point, namely the 
proper surface integral of the new 
$(8\pi)^{-1}N k|^{\rm ref}$, which agrees with 
the previous one in the $r \rightarrow \infty$ 
limit. As both $k|^{\rm ref}$'s are determined 
solely by the $B$ metric $\sigma_{ab}$, both 
(\ref{Hvalue}) and (\ref{newHvalue}) are Hamiltonian
boundary values as shown by Brown and York; and
we may view both expressions as Brown-York 
quasilocal energies
(``quasilocal'' as $B$ is finite before the $r \rightarrow
\infty$ limit is taken).\cite{BrownYork}

We demonstrate the asymptotic equivalence of 
(\ref{Hvalue}) and (\ref{newHvalue}) by comparing 
the leading-order terms in the separate asymptotic 
expansions for $k|^{\rm ref}$. Now, although the 
analysis below is certainly valid for case ii., 
in which $B$ tends to a round-sphere cut of future 
null infinity, this limit differs conceptually
from the other two. Hence, we postpone commenting 
on this case until the final section. We assume
that the $\Sigma$ Cauchy data obeys the required 
asymptotic conditions associated with
either an asymptotically flat solution\cite{York}
or an asymptotically anti-de Sitter solution\cite{HT}. 
In both cases, the physical $\Sigma$ three-metric
$h_{ij}$ approaches a fixed background metric $f_{ij}$. 
We need not consider the details of this approach, but 
we do note that in each case the fall-off conditions
on $h_{ij}$ imply that the  Ricci scalar 
of $B$ has the following asymptotic expansion:
\begin{equation}
{\cal R} \sim 2r^{-2}
+ {\cal R}^{3} r^{-3} {\,}.
\label{Rasymexpn}
\end{equation}
The ${\cal R}$ remainder need only be $o(r^{-3})$, 
i.~e.~it falls off {\em faster} than $r^{-3}$. 
Notice that the leading term is the scalar
curvature for a round sphere of radius $r$. (It is
often the case that the coefficient ${\cal R}^{3}$
actually vanishes identically, so after the leading
term the next non-zero term comes at a higher power
of inverse radius. Moreover, even if ${\cal R}^{3}$
does not vanish, a 
simple argument based on the Gau{\ss}-Bonnett
Theorem shows that it is a pure divergence on the
unit sphere.\cite{BLY1} 
These points do not affect the analysis here.) 
The expansion (\ref{Rasymexpn}) completely determines
the asymptotic behavior of the energy zero-points
we consider. Indeed, adopting (\ref{newreference})
in the asymptotically flat case, we compute 
\begin{equation}
       k|^{\rm ref} \sim 
      -2r^{-1} - {\textstyle \frac{1}{2}}{\cal R}^{3}
       r^{-2}{\,} ,
\label{asymexpn1}
\end{equation}
The leading term in (\ref{asymexpn1}) is exact for
a round sphere of radius $r$ embedded in $E^{3}$. 
Likewise, again defining $k|^{\rm ref}$ with 
(\ref{newreference}), in the  asymptotically 
anti-de Sitter case we get 
\begin{equation}
      k|^{\rm ref} \sim 
      -2/\ell - \ell r^{-2}  
      -{\textstyle \frac{1}{2}} 
       \ell {\cal R}^{3} r^{-3} {\,} .
\label{asymexpn2}
\end{equation}
In the next section, we show that with $k|^{\rm ref}$
constructed via the prescription listed in the
first paragraph, we obtain the exact same expansions
as (\ref{asymexpn1}) and (\ref{asymexpn2}). As this
prescription defines a Hamiltonian boundary value 
(\ref{Hvalue}) which determines the correct 
asymptotic energies, we are assured that 
(\ref{newHvalue}) does as well. We remark that upon
multiplication by $(8\pi)^{-1}N$ and subsequent 
integration over $B$, the leading term in 
(\ref{asymexpn1}) and the first two leading terms
in (\ref{asymexpn2}) diverge in the $r \rightarrow
\infty$ limit. These terms cancel corresponding divergent
terms which arise in the proper $B$ integral of the 
physical $(8\pi)^{-1}N k$.

As given in (\ref{newreference}), $k|^{\rm ref}$ 
is built solely with the intrinsic $B$ metric 
$\sigma_{ab}$; and, therefore, the surface integral
(\ref{newHvalue}) stands alone as a bona-fide Hamiltonian 
boundary value which determines the correct asymptotic
energies. Nevertheless, via an auxiliary isometric 
embedding of $B$ into a lightcone of the ambient 
reference spacetime, we can physically motivate the 
choice (\ref{newreference}). In the second section 
we discuss this motivation, the ``lightcone reference,'' 
in some detail.

\section{Static-time reference}

To follow the prescription outlined in the first
paragraph of the introduction, start with the
line-element for the reference spacetime $\referenceM$, 
\begin{equation}
      {\rm d}s^{2} = 
      -F(R){\rm d}t^{2} + F^{-1}(R){\rm d}R^{2} + 
      R^{2}({\rm d}\theta^{2} + \sin^{2}\theta 
      {\rm d}\phi^{2}) {\,} ,
\end{equation}
where $F(R) = 1$ for Minkowski spacetime and
$F(R) = (1 + R^{2}/\ell^{2})$ for an anti-de Sitter 
spacetime. In both cases $R$ is the round-sphere areal
radius. Now consider the surface $B$ embedded in a 
constant-$t$ slice $\referenceSigma$ associated with 
the above line element. Take $m^{a}$ and $\bar{m}^{a}$ 
as a complex null dyad tangent to $B$, and for 
notational ease let $\referencek_{ab}$ stand for 
$(k|^{\rm ref})_{ab}$, the extrinsic curvature of $B$ 
as embedded in $\referenceSigma$. Notice that 
$2\referencek_{m\bar{m}} := 
2\referencek_{ab} m^{a} \bar{m}^{b}$ and 
$\referencek_{mm} := \referencek_{ab} m^{a}m^{b}$ 
respectively capture the trace and trace-free
pieces of $\referencek_{ab}$. The embedding 
constraints satisfied by the intrinsic and extrinsic 
geometry of $B \subset \referenceSigma$ are 
the following:
\begin{eqnarray}
      R_{m\bar{m}m\bar{m}} & = &
           [\referencek_{m\bar{m}}]^{2} - 
      \referencek_{mm} \referencek_{\bar{m}\bar{m}} -
      {\textstyle \frac{1}{2}}{\cal R}
\label{1constraint}\\
      0 & = & \eth \referencek_{m\bar{m}} - 
      \bar{\eth} \referencek_{mm} {\,}.
\label{2constraint}
\end{eqnarray}
Here $\eth$ is the full $B$ ``eth'' operator and 
$R_{m\bar{m}m\bar{m}} := 
m^{i} \bar{m}^{j} m^{k}\bar{m}^{l}R_{ijkl}$, 
where $R_{ijkl}$ is the $\referenceSigma$ Riemann 
tensor\footnote{We use $(i,j,\cdots)$ as 
$\referenceSigma$ indices, whereas we have used 
$(a,b,\cdots)$ as $B$ indices.}. $R_{ijkl}$ vanishes 
for a static-time slice of Min\-kow\-ski spacetime, but
when  $\referenceSigma$ is a static-time slice of 
an anti-de Sitter spacetime $R_{m\bar{m}m\bar{m}}$
is non-zero. Indeed, were $B$ a round sphere of radius
$R = r$, we would have $R_{m\bar{m}m\bar{m}} = 1/\ell^{2}$.

	Let us first consider that case that 
$\referenceSigma$ is Euclidean three-space $E^{3}$.
In this case, were $B$ a round sphere of radius $R = r$, 
we would have $\referencek_{m\bar{m}} = -1/r$ and 
$\referencek_{mm} = 0$. Therefore, as $B$ tends to an 
infinite-area round sphere, we posit the expansions
\begin{eqnarray}
\referencek_{m\bar{m}} & \sim & -r^{-1} + 
\referencek^{2}_{m\bar{m}}
r^{-2}\\
\referencek_{mm} & = & O(r^{-2}) {\,}.
\end{eqnarray}
Plugging these expansions along with (\ref{Rasymexpn}) 
into (\ref{1constraint}), we can consistently match up
leading powers of $r$ to find the exact same expansion
as given in (\ref{asymexpn1}).

Now consider the other case when $\referenceSigma$ is 
a static slice of an anti-de Sitter spacetime. Then, 
were $B$ again a round
sphere of radius $R = r$, we would have $k_{mm} = 0$ and
\begin{equation}
\referencek_{m\bar{m}}
= - r^{-1}\sqrt{1 + r^{2}/\ell^{2}} {\,}.
\end{equation}
This suggests the {\em Ans\"{a}tze}
\begin{eqnarray}
\referencek_{m\bar{m}} & \sim & 
- 1/\ell - {\textstyle \frac{1}{2}}
\ell r^{-2} + \referencek_{m\bar{m}}^{3} r^{-3} \\
\referencek_{mm} & = & O(r^{-2}) {\,} .
\label{ansatz}
\end{eqnarray}
Moreover, we show below that with $m^{k}$ 
tangent to our slightly distorted sphere $B$,
\begin{equation}
R_{m\bar{m}m\bar{m}} = 1/\ell^{2} + 
O(r^{-4}) {\,} .
\label{curvefall1}
\end{equation}
With this result, (\ref{Rasymexpn}), and 
(\ref{ansatz}) plugged into (\ref{1constraint}), we 
consistently obtain the same expansion as given in 
(\ref{asymexpn2}). Notice that in both cases above
we have not used the
embedding constraint (\ref{2constraint}), although
we would need to were we also seeking the leading
asymptotics of $\referencek_{mm}$.

Let us verify equation (\ref{curvefall1}). 
Consider the natural foliation of $\referenceSigma$
and its associated triad
$\{p, q, \bar{q}\}$. Here $q^{k}$ is a 
complex dyad tangent to the round-sphere leaves of 
$\referenceSigma$, and $p^{k}$ is
the outward-pointing normal of a round sphere in
$\referenceSigma$. As $B$ is not perfectly round, 
at each point on $B$ we have an expansion of the 
form
\begin{equation}
       m^{k} = \alpha p^{k} + \bar{\beta} q^{k} +
                 \gamma \bar{q}^{k},
\end{equation}
with complex expansion coefficients $\alpha$, $\beta$, 
and $\gamma$. The coefficients must satisfy the 
orthogonality constraints 
$\alpha^{2} + 2\bar{\beta}\gamma = 0$ 
and 
$|\alpha |^{2} + |\beta |^{2} + |\gamma |^{2} = 1$.
Now as $B$ approaches a round sphere for large $r$, 
these constraints enforce the fall-off
conditions $\alpha = O(r^{-1})$, $\beta
= 1 + O(r^{-2})$, and $\gamma = O(r^{-2})$. 
Consider the explicit expressions for the components 
of the $\referenceSigma$ Riemann tensor
with respect to $\{p, q, \bar{q}\}$ 
(all are either zero or $\pm 1/\ell^{2}$, and in 
particular $R_{q\bar{q}q\bar{q}} = 1/\ell^{2}$), as well 
as the above orthogonality relationships between 
$\alpha$, $\beta$, 
and $\gamma$. Using these, we find that
\begin{equation}
R_{m\bar{m}m\bar{m}} = (|\beta|^{4} + 2|\alpha|^{2}
+ |\gamma|^{4} - 2 |\beta|^{2}|\gamma|^{2})/\ell^{2}{\,} .
\end{equation}
Finally, with the above fall-off for $\alpha$ and $\beta$ 
the second orthogonality constraint implies that 
$|\beta|^{4} + 2|\alpha|^{2}
= 1 + O(r^{-4})$, establishing (\ref{curvefall1}).

\section{Lightcone reference}

In this section we derive the choice (\ref{newreference}).
We begin by rewriting the $\referenceM$ line-element in 
terms of an outgoing null coordinate 
$w := t - \ell\arctan(R/\ell)$, thereby obtaining
\begin{equation} 
      {\rm d}s^{2} = 
      - F(R) {\rm d}w^{2} - 2{\rm d}w {\rm d}R
      + R^{2}({\rm d}\theta^{2} + \sin^{2}\theta 
        {\rm d}\phi^{2}){\,} 
\end{equation}
for which constant-$w$ 
surfaces are outgoing null cones. Pick one such 
cone $\referenceN$ and embed $B$ in it. One can imagine a 
spherical cross section of $\referenceN$ which is ``pushed up
and down'' the null generators of $\referenceN$ in an
angle-dependent fashion. Under such a transformation the 
metric of the cross section undergoes a conformal 
transformation to become isometric with $B$. We then work 
with the chain of inclusions $B \subset \referenceN \subset
\referenceM$, and further consider a three-slice $\referenceSigma$
which spans $B$. As $B$ is not perfectly round, 
$\referenceSigma$ is not a static slice of $\referenceM$.
Further, in general $\referenceSigma$ will not be a moment of 
time symmetry, and hence will have a non-vanishing extrinsic
curvature tensor $\referenceK_{ij}$. Projection into $B$ of 
the free indices of $\referenceK_{ij}$ defines an extrinsic
curvature tensor $\referencel{}_{ab}$ on $B$. Now, since $B$
is realized as a 2-surface in $\referenceM$, the embedding 
$B \subset \referenceM$ must satisfy the following 
constraint:
\begin{equation}
       (\referencek){}^{2}
     - \referencek{}^{ab}\referencek{}_{ab} -
       (\referencel){}^{2}
     + \referencel{}^{ab}
       \referencel{}_{ab} - {\cal R}  =  
       2\Re_{m\bar{m}m\bar{m}} {\,},
\label{embeddingconstraint}
\end{equation}
where $\Re_{m\bar{m}m\bar{m}} := 
\Re_{\mu\nu\lambda\sigma} 
m^{\mu}\bar{m}^{\nu} m^{\lambda} \bar{m}^{\sigma}$
denotes projection of the $\referenceM$ Riemann tensor 
$\Re_{\mu\nu\lambda\sigma}$ into $B$ (again $m^{\mu}$
is the complex null dyad tangent to $B$).\footnote{With
$(\mu,\nu,\cdots)$ as $\referenceM$ indices.}
There are of course other constraint equations associated
with the embedding $B \subset \referenceM$, but we need 
not consider them here.

To construct the lightcone reference,
first note that the geodetic congruence $\referenceN$ is
sheer-free,
which means that the complex shear
$\referencesigma := \referencek_{mm} + \referencel_{mm}$
of $\referenceN$ vanishes. 
The vanishing of $\referencesigma$ thus implies that the
trace-free piece of $\referencek_{ab}$ equals minus the
trace-free piece of $\referencel_{ab}$; whence
Eq.~(\ref{embeddingconstraint}) becomes
\begin{equation}
       (\referencek){}^{2}
     - (\referencel){}^{2}
     - 2 {\cal R} = 
     4 \Re_{m\bar{m}m\bar{m}} {\,} .
\label{tracesgone}
\end{equation}
Now, by performing a simple coordinate transformation
on the above line-element, one can quickly argue that 
\begin{equation}
\Re_{m\bar{m}m\bar{m}} = 
1/\ell^{2}
\label{curvefall2}
\end{equation}
even for our distorted surface $B$ (this is also
the result were $B$ perfectly round).
Of course, the projection (\ref{curvefall2})
vanishes straightaway if $\referenceM$ is Minkowski 
spacetime. Next, we choose $\referenceSigma$ by the 
condition $\referencel = 0$. This sets to zero 
the component of gravitational momentum normal to $B$,
defining $\referenceSigma$ as the ``rest frame'' 
associated with $B$.\cite{Lau} With 
$\referencel = 0$, we can immediately solve 
(\ref{tracesgone}) for $\referencek$ to get 
(\ref{newreference}). Notice that the 
$\referenceSigma$ of this 
section would also be a static-time slice, were $B$ a 
round sphere (in $\referenceM$ static-time slices
intersect lightcones in round spheres). 
Therefore, as $B$ tends to a round
sphere in the $r \rightarrow \infty$ limit, it
is not surprising that the static-time and lightcone 
references determine the same asymptotic energies.

\section{Null infinity limit}

In closing, let us return to case ii., the 
Trautman-Bondi-Sachs
energy associated with null infinity for asymptotically
flat spacetimes. In this case, as well as being a surface
drawn in $\Sigma$, $B$ is also a cut of an
outgoing null congruence $N \subset M$, and whether
$r$ is an areal or affine radius we have an expansion
for the $B$ curvature scalar which matches 
(\ref{Rasymexpn}).\cite{BLY1} Then for either choice of 
zero-point we determine an expansion for $k|^{\rm ref}$ of the
form (\ref{asymexpn1}). Therefore, (\ref{newHvalue}) has 
the correct form to
give the Trautman-Bondi-Sachs energy in the limit that
$B$ becomes a round two-sphere cut of null 
infinity.\footnote{This was also noted in \cite{BLY2}.}

\section{Acknowledgments}  For helpful discussions I thank 
A.~Anderson, J.~D.~Brown, N.~\'{O} Murchadha, and J.~W.~York. 
This work as been supported by the National Science Foundation 
(NSF grant No.~PHY-9413207 to the University of North Carolina).


\begin{thebibliography}{99}

\bibitem{ADM} R.~Arnowitt, S.~Deser, and C.~Misner,
              in {\em Gravitation: an Introduction
              to current research} (Wiley, NewYork, 1962).

\bibitem{TBS} A.~Trautman, Bull.~Acad.~Pol.~Sci.~S\'{e}rie
              des Sci.~Math.~Ast.~Phys.~{\bf 6}, 407 (1958);
              H.~Bondi, M.~G.~J.~van der Burg, and
              A.~W.~K.~Metzner, Proc.~R.~Soc.~London A{\bf 269},
              21 (1962); R.~K.~Sachs, Proc.~R.~Soc.~London
              A{\bf 270}, 103 (1962); R.~K.~Sachs, 
              Phys.~Rev.~{\bf 128}, 2851 (1962).

\bibitem{AD}  L.~Abbott and S.~Deser, Nucl.~Phys.~B {\bf 195}, 
              76 (1982).

\bibitem{HawkingEllis} S.~W.~Hawking and G.~R.~Ellis,
                       {\em The Large Scale Structure of
                       Space-time} (Cambridge University  
                       Press, Cambridge, 1973).   
               
\bibitem{BrownYork} H.~Braden, J.~D.~Brown, B.~Whiting,
                    and J.~W.~York, Phys.~Rev.~D {\bf 42},
                    3376 (1990); J.~D.~Brown and J.~W.~York,
                    Phys.~Rev.~D {\bf 47}, 1407 (1993).

\bibitem{BLY1} J.~D.~Brown, S.~R.~Lau, and J.~W.~York,
               Phys.~Rev.~D {\bf 55} 1977 (1997).


\bibitem{Brown} J.~D.~Brown, J.~Creighton, and R.~B.~Mann,
                Phys.~Rev.~D {\bf 50}, 6394 (1994).

\bibitem{HawkingHorowitz} S.~W.~Hawking and G.~T.~Horowitz,
                          Class.~Quantum Grav.{\bf 13}, 1487 
                          (1996).

\bibitem{York} J.~W.~York in {\em Sources of Gravitational
               Radiation}, edited by L.~Smarr (Cambridge
               University Press, Cambridge 1979).

\bibitem{HT} M.~Henneaux and C.~Teitleboim, 
             Commun.~Math.~Phys {\bf 98}, 391 (1985).

\bibitem{Lau} S.~R.~Lau, Class. Quantum Grav. {\bf 13}, 
              1509 (1996).

\bibitem{BLY2} J.~D.~Brown, S.~R.~Lau, and J.~W.~York,
               Phys.~Rev.~D {\bf 59}, 064028 (1999).
\end{thebibliography}
\end{document}